\documentclass[journal,twoside,web]{ieeecolor}
\usepackage{generic}
\usepackage{cite}
\usepackage{amsmath,amssymb,amsfonts}
\usepackage{algorithmic}
\usepackage{graphicx}
\usepackage{amsmath}
\usepackage{algorithm,algorithmic}
\usepackage{hyperref}
\usepackage{multirow}
\usepackage{booktabs}
\usepackage{amssymb}
\usepackage{tabularx}
\usepackage{pifont}
\hypersetup{hidelinks=true}
\usepackage{textcomp}
\usepackage[utf8]{inputenc}
\usepackage{booktabs}       
\usepackage{tabularx}       
\usepackage{siunitx}        
\usepackage{threeparttable} 
\def\BibTeX{{\rm B\kern-.05em{\sc i\kern-.025em b}\kern-.08em
    T\kern-.1667em\lower.7ex\hbox{E}\kern-.125emX}}
\begin{document}
\title{AtlasSeg: Atlas Prior Guided Dual-U-Net for Tissue Segmentation in Fetal Brain MRI}
\author{Haoan Xu\footnotemark*, Tianshu Zheng\footnotemark*, Xinyi Xu, Yao Shen, Jiwei Sun, Cong Sun, Guangbin Wang, Zhaopeng Cui, and Dan Wu
\thanks{Manuscript submitted 10 March 2025. This study was supported by the National Natural Science Foundation of China (82122032), and Science and Technology Department of Zhejiang Province (2022C03057, 202006140). (Corresponding author: Dan Wu)}
\thanks{Haoan Xu, Tianshu Zheng, Xinyi Xu, Yao Shen, Jiwei Sun and Dan Wu are with Key Laboratory for Biomedical Engineering of Ministry of Education, Department of Biomedical Engineering, College of Biomedical Engineering \& Instrument Science, Zhejiang University, Hangzhou, 310027, China (email: haoanxu@zju.edu.cn; zhengtianshu@zju.edu.cn; xuxinyi\_bme@zju.edu.cn; shenyao.bme@ zju.edu.cn; 12118660@zju.edu.cn; danwu.bme@zju.edu.cn).}
\thanks{Cong Sun is with Department of Radiology, Beijing Hospital, National Center of Gerontology, Institute of Geriatric Medicine, Chinese Academy of Medical Sciences, Beijing, 100730, China. (email: suncong\_jida1991@163.com)}
\thanks{Guangbin Wang is with Department of Radiology, Shandong Provincial Hospital Affiliated to Shandong First Medical University, Jinan, Shandong, 276899, China. (email: wgb7932596@hotmail.com)}
\thanks{Zhaopeng Cui is with State Key Lab of CAD\&CG, Zhejiang University, Hangzhou, China. (email: zhpcui@zju.edu.cn)}}

\maketitle
\footnotetext[1]{These authors contributed equally to this work.}

\begin{abstract}
Objective: Accurate automatic tissue segmentation in fetal brain MRI is a crucial step in clinical diagnosis but remains challenging, particularly due to the dynamically changing anatomy and tissue contrast during fetal development. Existing segmentation networks can only implicitly learn age-related features, leading to a decline in accuracy at extreme early or late gestational ages (GAs). Methods: To improve segmentation performance throughout gestation, we introduce AtlasSeg, a dual-U-shape convolution network that explicitly integrates GA-specific information as guidance. By providing a publicly available fetal brain atlas with segmentation labels corresponding to relevant GAs, AtlasSeg effectively extracts age-specific patterns in the atlas branch and generates precise tissue segmentation in the segmentation branch. Multi-scale spatial attention feature fusions are constructed during both encoding and decoding stages to enhance feature flow and facilitate better information interactions between two branches. Results: We compared AtlasSeg with six well-established networks in a seven-tissue segmentation task, achieving the highest average Dice similarity coefficient of 0.91. The improvement was particularly evident in extreme early or late GA cases, where training data was scare. Furthermore, AtlasSeg exhibited minimal performance degradation on low-quality images with contrast changes and noise, attributed to its anatomical shape priors. Conclusion: AtlasSeg demonstrated enhanced segmentation accuracy, better consistency across fetal ages, and robustness to perturbations, making it a powerful tool for reliable fetal brain MRI tissue segmentation, particularly suited for diagnostic assessments during early gestation.
\end{abstract}

\begin{IEEEkeywords}
Fetal brain MRI; tissue segmentation; fetal brain atlas; convolutional neural network.
\end{IEEEkeywords}

\section{Introduction}
\label{sec:introduction}
\IEEEPARstart{M}{agnetic} resonance imaging (MRI) of the fetal brain has become an essential tool in prenatal studies, significantly improving the understanding of fetal growth and development \cite{ertl-wagner_fetal_2002}, \cite{griffiths_use_2017}. It plays an important role in clinical examination of prenatal brain disorders due to its superior image resolution and diverse tissue contrasts, which are considered more informative than ultrasonography \cite{davidson_fetal_2021}, \cite{aughwane_placental_2020}. The advancements in fetal MRI were attributed to the development of fast imaging techniques \cite{ferrante_slice--volume_2017}, improved motion correction and advanced reconstruction algorithms \cite{gholipour_robust_2010}, \cite{kainz_fast_2015}. These technological improvements enabled the generation of detailed, high-resolution three-dimensional volumes of the fetal brain, providing benefits for both diagnostic and quantitative assessments \cite{ebner_automated_2020}.

\begin{figure*}[!t]
\centering
\includegraphics[width=\textwidth]{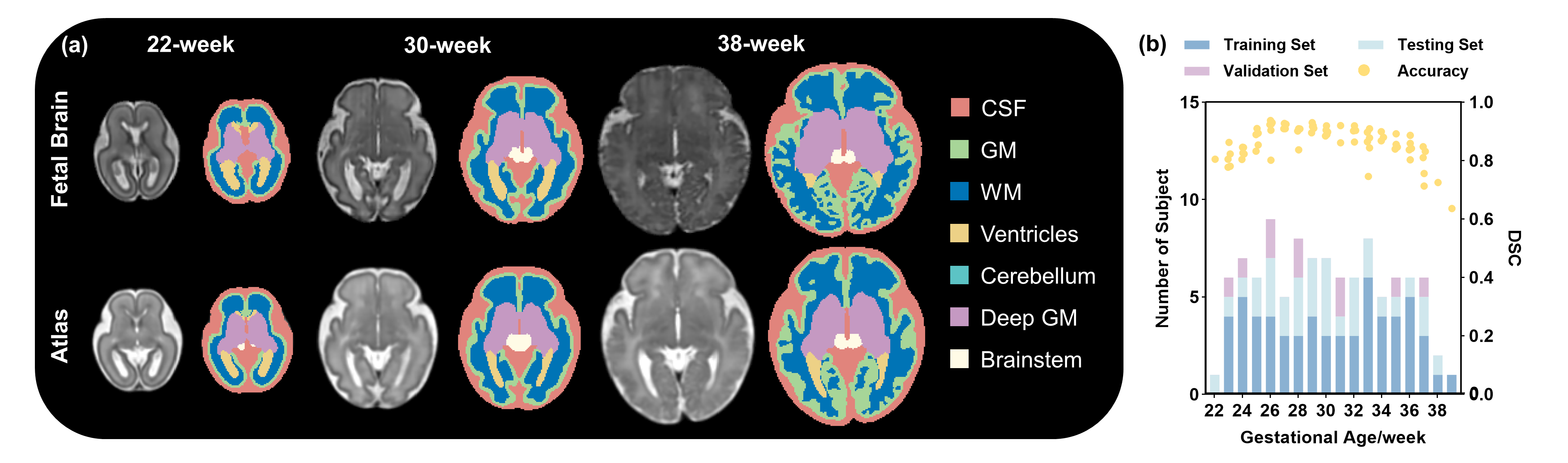}
\centering
\caption{(a) Example of the fetal brains, atlases, and corresponding tissue maps at younger, middle, and older ages. (b) Imbalanced distribution of fetal subjects and segmentation accuracy (yellow dots, from our previous work \cite{xu_site_2024}) across GA, with reduced performance in the youngest and oldest fetal brains.}
\label{fig1}
\end{figure*}

Central to the quantification of fetal brain anatomy is tissue segmentation \cite{makropoulos_review_2018}. As shown in Fig. \ref{fig1}a,  fetal brain is typically categorized into seven structures: cerebrospinal fluid (CSF), gray matter (GM), white matter (WM), ventricles, cerebellum, deep gray matter (dGM), and brainstem. Many disease or development indicators can be generated from these segmentation labels. For example, cortical volume can be calculated based on the GM label, which is useful for diagnosing congenital heart disease \cite{sadhwani_fetal_2022}, \cite{clouchoux_delayed_2013}. However, segmentation of fetal brain is vastly different from that of the adult brains, due to the thin miniature sized fetal brain, weak contrast, and dynamically changing anatomy and tissue contrast of the fetal brains. This field has evolved from manual or semi-automatic approaches, such as multi-atlas registration techniques \cite{hutchison_atlas-based_2008}, to more advanced deep learning (DL) solutions. Most DL-based methods train U-shape convolutional neural networks (CNNs) with manually labeled ground truths in an end-to-end manner, such as the nnU-Net \cite{isensee_nnu-net_2021}, which has become the state-of-the-art (SOTA) approaches in segmentation of the fetal brain \cite{siddique_u-net_2021}.

Despite these efforts, segmentation accuracy of the fetal brain remains limited compared to that in the adult brains \cite{ciceri_review_2023}. We think one primary reason is associated with the rapidly changing fetal brain with gestational age (GA), as shown by the examples in the top row of Fig. \ref{fig1}a. This variability poses a major obstacle for the existing end-to-end networks, which can only implicitly learn GA-related information. As a result, the network's adaptation to the wide range of anatomical variations across different GA is limited. Furthermore, another issue is the imbalanced distribution of GA in the training data (Fig. \ref{fig1}b), as less patients are scanned at early GA or very late GA close to delivery. This uneven distribution can adversely affect the network's learning, making it less effective in segmenting fetal brain tissues at early or late stages of pregnancy. Because fetal brain atlases from mid-to-late gestations have been generated \cite{xu_spatiotemporal_2022}, \cite{gholipour_normative_2017}, \cite{serag_construction_2012} and made publicly available, we hypothesized that by borrowing the anatomical information of the atlases at corresponding GA weeks, we might be able to improve the segmentation accuracy.

\subsection{Related Work}
In early studies, automatic tissue segmentation of 3D fetal brain MRI often relied on structural information provided by atlases \cite{oishi_multi-contrast_2011}, \cite{habas_spatiotemporal_2010}. These multi-atlas segmentation (MAS) techniques usually involved registering the unlabeled source volume to several target atlases and then back-transforming atlas labels and fusing of the labels to obtain segmentation \cite{gholipour_normative_2017}. Traditional registration methods could be replaced by some DL-based deformable registration frameworks \cite{li_coupling_2022} like VoxelMorph \cite{balakrishnan_voxelmorph_2019}, for improved accuracy and faster segmentation. However, manual correction remained necessary after MAS \cite{vasung_spatiotemporal_2020}.

Recently, U-Net \cite{ronneberger_u-net_2015} had emerged as a benchmark in medical image segmentation. Most DL-based fetal brain MRI segmentation networks utilized U-Net or its variants \cite{ciceri_review_2023}. Payette et al. \cite{payette_fetal_2023} used a U-Net-based architecture to segment fetal brain ventricles for longitudinal and volumetric analyses. Zhao et al. \cite{zhao_automated_2022} used a 3D extension of the original 2D U-Net to segment the whole-brain volume into six tissues. Various enhancements, such as dense skip pathways  \cite{stoyanov_unet_2018} and attention mechanisms \cite{oktay_attention_2018}, have been used to strength the performance of U-Net. Dou et al. \cite{dou_deep_2021} proposed an attentive FCN for fetal cortex segmentation, which integrated a backbone U-Net with stage-wise attention refinements at each decoder level. To make full use of noisy labels, Fetit et al. \cite{fetit_deep_nodate} proposed a human-in-the-loop method where the annotator could refine the labels obtained from MAS to segment cortex. Karimi et al. \cite{karimi_learning_2023} introduced a novel training approach with label-smoothing procedure and certainty-adapted loss function to segment fetal brain tissue from MAS-based noisy labels. Xu et al. \cite{xu_efficient_2024} investigated the relationship between fetal brain MRI segmentation and physical resolution, who found that downsampling the images to the physical resolution (about 2mm) did not result in a significant decrease in accuracy, which could help in the development of faster and more efficient segmentation methods.

Other studies replaced parts of U-Net with transformer \cite{vaswani_attention_nodate} modules. Huang et al. \cite{huang_deep_2023} proposed a lightweight model that integrated a CNN-based encoder-decoder design with a contextual transformer module. Qi et al. \cite{qi_mg-net_2024} replaced the bottleneck with a multi-scale deformable transformer and incorporated graph convolution-based attention into decoder, achieving more precise segmentation. Pecco et al. \cite{pecco_optimizing_2024} tested the performance of U-Net, U-Net transformer (UNETR) \cite{hatamizadeh_unetr_2022}, and SwinUNETR \cite{crimi_swin_2022} on an internal dataset of 172 fetuses and an external dataset of 131 fetuses. Validation on the internal dataset showed that transformer-based models performed worse than CNN-based models, with SwinUNETR consistently outperforming UNETR. External validation showed comparable performance between SwinUNETR and U-Net. Furthermore, SwinUNETR demonstrated better performance than U-Net during the late-fetal period, but performed worse during the mid-fetal period. 

\subsection{Contributions}

\begin{figure*}[!t]
\centering
\includegraphics[width=\textwidth]{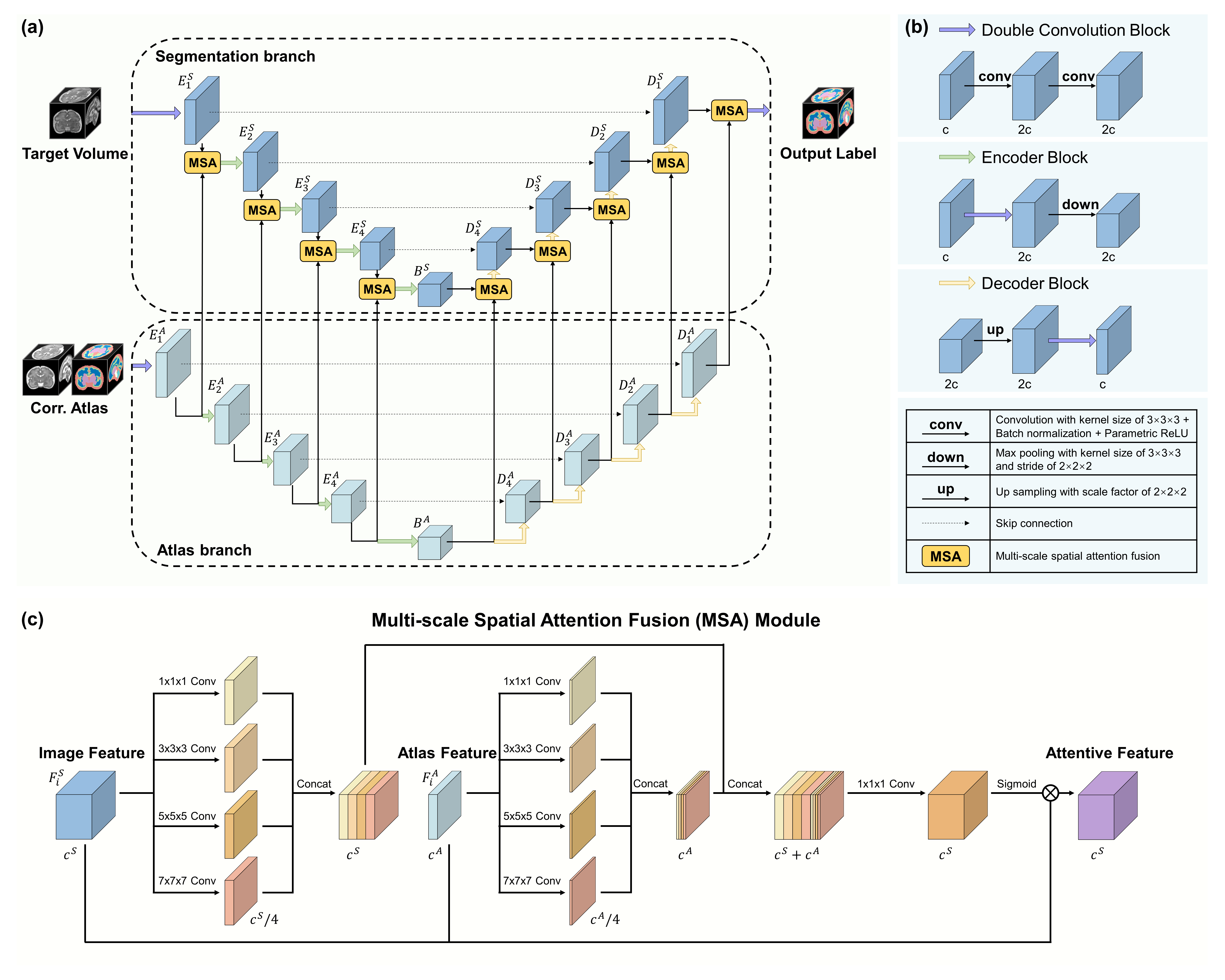}
\centering
\caption{(a) Architecture of AtlasSeg, which is built upon two parallel 3D U-Net backbones with encoder-decoder structures and skip connections. Dense multi-scale spatial attention fusion modules interlink two branches to build a dense feature flow. The network takes three 96×96×96 patches as input. (b) Convolutional blocks and operations used in (a). (c) Proposed MSA. This fusion module took image and atlas features from corresponding branches as inputs, and outputs fused feature for further segmentation. A group of convolutions with kernel sizes of [7,5,3,1] were used to extract multi-scale features.}
\label{fig2}
\end{figure*}

To improve the segmentation of fetal brain MRI across various GAs, and to address the variability in anatomical structures observed during fetal development, we proposed an Atlas prior guided Segmentation network (AtlasSeg). The proposed AtlasSeg integrated GA prior information, as encoded in the corresponding atlas to assist the segmentation. AtlasSeg consisted a dual-U-Net architecture, simultaneously processing the MRI volume and corresponding atlas image and labels at the matching GA. Stage-wise dense attention connections were constructed at different convolution stages for feature fusion to enhance the flow of features between the two branches. We compared the performance of AtlasSeg against six well-known networks: 3D U-Net \cite{ronneberger_u-net_2015}, U-Net++ \cite{stoyanov_unet_2018}, Attention U-Net (AttU-Net) \cite{oktay_attention_2018}, MixAttNet \cite{dou_deep_2021}, UNETR \cite{hatamizadeh_unetr_2022}, and SwinUNETR \cite{crimi_swin_2022}. We divided the pregnancy period into four subgroups, and the quantitative results showed that the performance of AtlasSeg was the most accurate in each group and more robust across groups. Furthermore, through contrast- and noise-corrupted image tests, we found the anatomical shape priors embedded in AtlasSeg enhanced its generalizability against such low-quality perturbations.

The paper is organized as follows: the model architecture and attention mechanism are discussed in Section II. The experiment details and the introduction of fetal MRI dataset are described in Section III. Section IV describes the experimental results, followed by a discussion in Section V and a conclusion in Section VI. The relevant code has been released on GitHub (https://github.com/zjuwulab/AtlasSeg).

\section{Method}
\subsection{Model Architecture}

Fig. \ref{fig2} illustrates the architecture of our proposed AtlasSeg for fetal brain MRI tissue segmentation. AtlasSeg utilizes two parallel U-Net structures, each featuring an encoder-decoder design and skip connections at every convolutional stage. These twin U-Nets serve distinct roles, one as the segmentation branch for processing the target fetal brain MRI volume, while the other as the atlas branch, handling the corresponding atlas comprising two channels of image and labels. To ensure efficient information exchange, dense attentive connection modules are implemented between the dual U-Nets. The similar structure of two branches facilitates the establishment of attention connections, as corresponding feature maps are of the same size.

The input MRI volume first passes through a double convolution block (shown in Fig. \ref{fig2}b), comprising two convolution operations, each followed by batch normalization (BN) \cite{ioffe_batch_2015} and Parametric Rectified Linear Unit (PReLU) \cite{he_delving_2015} for activation. The initial feature map in the segmentation branch's encoder, denoted as $E_1^S$, captures the semantic features of the target tissue. Simultaneously, in the atlas branch, the corresponding atlas image and labels are concatenated into a two-channel input and subjected to an identical double convolution block with separate parameters, resulting in $E_1^A$. Both $E_1^S$ and $E_1^A$ are subsequently fed into an attentive fusion module, which will be discussed in the following section. The output of fusion is further processed by the encoder block in Fig. \ref{fig2}b, consisting of a double convolution block and a downsampling block. After encoding, the bottleneck feature maps, $B^S$ and $B^A$, passes through a fusion module and are then decoded via a block containing a transposed convolution-based upsampling and another double convolution block. In the final decoding stage, the output of segmentation branch undergoes convolution with 1*1*1 kernel size and 8 output channels, followed by a softmax activation function to generate the segmentation labels.

Segmentation in AtlasSeg can be performed on entire volume or based on patches. In our implementation, we choose to use patches of size $96^3$. Each encoding/decoding block contains either a downsampling or upsampling operation, leading to progressively reduced feature map sizes of $48^3$, $24^3$, $12^3$, $6^3$. As shown in Fig. \ref{fig2}b, the channel number varies during the encoding and decoding phases. In the segmentation branch, the initial channel number $c^S$ is set at 32, which sequentially expands to 64, 128, 256, and 512 channels at each respective stage. Conversely, the atlas branch starts with a lower initial channel number $c^A$ of 8, according to our experimental result that will be discussed in the ablation tests.

\subsection{Multi-scale Attention Atlas Fusion Module}

As shown in Fig. \ref{fig2}c, multi-scale spatial attention (MSA) feature fusion module is used to fuse features from segmentation and atlas branches at corresponding stage i, which are denoted as $F_i^S$ and $F_i^A$, respectively. This setup allows the module to dynamically weight the features from segmentation branch and prioritize anatomical details and contextual information guided by atlases. Due to the need of feature fusions at different stages of the network, spatial attention (SA) with just a 3×3×3 convolution is insufficient to provide enough receptive field. Therefore, MSA is used to broaden the receptive field of the attention map and capture contextual information at different spatial resolutions.

In MSA, the feature map of segmentation branch $F_i^S$ with channel number of $c^S$ is processed by a group of convolutions with kernel sizes of [1, 3, 5, 7] and output channel number of $c^S/4$, follow by BN and PReLU. The atlas feature $F_i^A$ undergoes a parallel process, albeit with an output channel number of $c^A/4$. All attention-enhanced features are concatenated together to a channel number of $c^S+c^A$, then pass through a 1×1×1 convolution for merging, followed by sigmoid activation to produce the attention weighting map $A$. The attention weighting map $A$ are element-wise multiplied with $F_i^S$ to generate the deep fusion output. The overall operation of MSA can be summarized by the following form:
\begin{align}
F_i' &= A \otimes (F_i^S) = \sigma \left( f(F_i^S; \theta_i^S) \oplus f(F_i^A; \theta_i^A) \right) \otimes (F_i^S),
\label{eq1}
\end{align}
where $\oplus$ represents channel-wise concatenation, and $\otimes$ is element-wise multiplication, $\sigma$ represent the 1*1*1 convolution and sigmoid activation, $f$ represents MSA, $\theta_i^S$ and $\theta_i^A$ denote the parameters of convolutional groups.

\subsection{Loss Function}
We trained our network with the unweighted sum of cross-entropy (CE) loss and dice loss \cite{milletari_v-net_2016}. $p_{v,i}$ and $g_{v,i}$ denotes the predicted label and ground truth at voxel $v$ for class $i$. $V$ is the number of voxels. $l$ is the number of classes. CE loss is defined as:
\begin{equation}
L_{\text{CE}} = - \sum_{i=1}^{l} \sum_{v=1}^{V} \left[ g_{v,i} \log(p_{v,i}) + (1 - g_{v,i}) \log(1 - p_{v,i}) \right],
\label{eq2}
\end{equation}
and dice loss is defined as:
\label{eq3}
\begin{equation}
L_{\text{Dice}} = 1 - \sum_{i=1}^{l} \frac{2 \sum_{v=1}^{V} g_{v,i} p_{v,i}}{\sum_{v=1}^{V} (g_{v,i})^2 + \sum_{v=1}^{V} (p_{v,i})^2}.
\end{equation}
The final loss $L$ is defined as:
\begin{equation}
L = L_{\text{CE}} + L_{\text{Dice}}.
\label{eq4}
\end{equation}

\section{Experiment}
\subsection{Data acquisition and preprocessing}
A total of 180 fetal brain MRIs in at least three orthogonal orientations (axial, coronal, and sagittal) were collected under Institutional Review Board approval. The data were acquired on a 3T Siemens Skyra scanner (Siemens Healthineers, Erlangen, Germany) with an abdominal coil, using a T2-weighted half-Fourier single-shot turbo spin-echo (HASTE) sequence with the following protocol: repetition time/echo time = 800/97 ms, in-plane resolution = 1.09 × 1.09 mm, field of few = 256 × 200 mm, thickness = 2 mm, partial Fourier factor = 5/8, echo train length = 102, and GRAPPA factor = 2, number of calibration line = 42 with an interleaved acquisition. Informed consent was obtained from all pregnant volunteers participating in the study.

31 cases were excluded before data preprocessing due to low image quality, such as low signal-to-noise ratio, signal voids, and severe fetal motion. The remaining 149 cases were processed with the following pipeline. First, bias field correction was performed using the N4 algorithm \cite{tustison_n4itk_2010}. Then, fetal brains were extracted using a CNN-based brain extraction tool \cite{ebner_automated_2020}, followed by 3D non-local means denoising \cite{coupe_optimized_2008}. Each stack was normalized to its maximum intensity. Next, motion correction and super resolution reconstruction were executed utilizing the NiftyMIC toolkit \cite{ebner_automated_2020} to generate isotropic volumes at a resolution of 0.8 mm. 47 cases were further excluded due to motion artifacts or poor image quality after reconstruction, resulting in 102 subjects (GA: 22.4-39.0 weeks) for further segmentation. The remaining fetal brain volumes were rigidly registered to the spatiotemporal fetal brain atlas \cite{serag_construction_2012} using FLIRT \cite{jenkinson_fsl_2012}. The registered volumes were resized to dimensions of 192×192×144 with zero-padding. 

To generate ground truth tissue labels, all 3D reconstructed fetal brain volumes were segmented using an automatic MAS method provided by DrawEM. After MAS, the volumes were categorized into 8 classes: CSF, GM, WM, ventricles, cerebellum, dGM, brainstem, and background. The coarse segmentations underwent several rounds of manual corrections by three raters over the span of a year to ensure accuracy. The supporting atlas images and labels utilized in the network were derived from the 4D spatiotemporal fetal brain atlas from 23 to 38 weeks of GA, established by Xu et al. \cite{xu_spatiotemporal_2022}. The GA of input MRI was rounded to the nearest GA and then paired with the corresponding atlas. For images outside the atlas range, such as those at 39 weeks of gestation, the closest available atlas was used. 

\subsection{Evaluation Metrics}
To quantitatively assess the performance of different networks, we employed three evaluation metrics: Dice Similarity Coefficient (DSC), 95 percent Hausdorff Distance (95HD), and Average Symmetric Surface Distance (ASSD). DSC is one of the most common metrics in medical segmentation \cite{liu_review_2021}, which is region-based and measures the spatial overlap between the ground truth and the predicted segmentation. 95HD and ASSD are both boundary-based metrics. HD calculates the distance of a set to the nearest point in the other set, which is used to measure the similarity between two boundaries:
\begin{equation}
\mathrm{HD}(G, P) = \max \left\{
\begin{aligned}
& \mathrm{sup}_{g \in S_G} \mathrm{inf}_{p \in S_P} d(g, p), \\
& \mathrm{sup}_{p \in S_P} \mathrm{inf}_{g \in S_G} d(p, g)
\end{aligned}
\right\},
\label{eq5}
\end{equation}
where $S_G$ and $S_P$ represents the sets of surface points in the ground truth $G$ and the predicted segmentation $P$, $d(g,p)$ represents the Euclidean distance between points $g$ and $p$. 95HD considers the distance below 95\% of all points, which makes it more robust to outliers compared to the standard HD. ASSD assesses the average distance from points on the predicted segmentation surface to the ground truth: 
\begin{equation}
\mathrm{ASSD}(G, P) = \frac{1}{|S_G| + |S_P|} \left[
\begin{aligned}
& \sum_{g \in S_G} \min_{p \in S_P} d(g, p) \\
& + \sum_{p \in S_P} \min_{g \in S_G} d(p, g)
\end{aligned}
\right].
\end{equation}

\subsection{Algorithm Comparison}
We compared the performance of AtlasSeg with six well-known models: 3D U-Net \cite{ronneberger_u-net_2015}, U-Net++ \cite{stoyanov_unet_2018}, AttU-Net \cite{oktay_attention_2018}, MixAttNet \cite{dou_deep_2021}, UNETR \cite{hatamizadeh_unetr_2022}, and SwinUNETR \cite{crimi_swin_2022}. U-Net++ is an advanced version of U-Net, which introduced nested, dense skip pathways between encoder and decoder. AttU-Net enhanced U-Net with attention gates into skip connections, which achieved remarkable performance in many medical image tasks. MixAttNet was specifically designed for fetal brain MRI cortical plate segmentation. It combined a U-Net backbone with multi-scale attention refinement modules, which had been proven to be effective in segmenting the fetal brain cortex. Both UNTER and SwinUNETR were transformer-based networks. UNETR built upon U-Net by utilizing a hierarchical transformer-based encoder to process input patches, enabling global contextual reasoning through self-attention mechanisms. SwinUNETR extended UNETR with the shifted window mechanism, which divided the input image into non-overlapping windows for efficiency. All networks were trained from scratch under the same schemes and hyper-parameters. 

\subsection{Generalizability study}
This section aims to evaluate the generalizability of different networks when inferring low-quality images. The networks were trained on the original images and tested on three types of low-quality images with added perturbations. The first two perturbations were designed to assess the robustness to varying input contrasts. We employed gamma correction to modify the image contrast, where the image was transformed as $I=I^\gamma$, followed by normalization. When $\gamma$ was less than 1, the images were visually brighter after normalization. When $\gamma$ was greater than 1, the images appeared darker after normalization. The values of $\gamma$ were chosen from the ranges [0.5, 0.6, 0.7, 0.8, 0.9] and [1.1, 1.2, 1.3, 1.4, 1.5], respectively. The third test assessed the networks’ generalizability to noise. Gaussian noise with a mean of 0 and standard deviations (SD) in the range [0.01, 0.02, 0.03, 0.04, 0.05] was added to the images, followed by normalization.

\subsection{Ablation Study}
We conducted ablation studies to validate the effectiveness of atlas branch and different feature fusions. In the first ablation, we evaluated the performance of adding the atlas branch into the U-Net backbone using late attention, which simply concatenates the feature maps from the Atlas branch into the segmentation branch. We then introduced SA and MSA to generate spatial attention maps to facilitate feature fusion. Additionally, we assessed the impact of adding MSA directly into backbone U-Net. This was shown as the segmentation branch in Figure 2a. The second ablation investigated the impact of the position of feature fusion within the network, including 1) Early fusion, with attention at the initial stage $E_1^S$; 2) Late fusion, with attention at the final stage $D_1^S$; 3) Encoder fusion, with attention at encoder stages; 4) Decoder fusion, with attention at decoder stages; and 5) the full AtlasSeg model, which integrated feature fusion at every stage. In the last ablation, with keeping the channel number of segmentation branch fixed at 32, we tested the channel number of atlas branch of [4,8,16,32].

\begin{table*}
\centering
\caption{Comparison of inference results using different methods (n=32), averaged across all labels. The best value for each measurement has been shown in boldface text. Except for 95HD of U-Net++ and AttU-Net, other measurements show significant difference (p<0.001) compared to AtlasSeg using paired t-test.}
\label{table1}
\fontsize{9pt}{14pt}\selectfont
\resizebox{13cm}{!}{
\begin{tabularx}{13cm}{@{\hskip 0.15in}X@{\hskip 0.1in}X@{\hskip 0.2in}X@{\hskip 0.2in}X@{\hskip 0.1in}}
\hline
Network          & DSC{\scalebox{1}{$\uparrow$}}              & 95HD(mm){\scalebox{1}{$\downarrow$}}       & ASSD(mm){\scalebox{1}{$\downarrow$}}           \\ 
\hline
U-Net            & 0.8891±0.0202  & 7.0527±0.6240 & 2.0773±0.2331 \\
U-Net++          & 0.8957±0.0174  & 5.1044±0.8766 & 1.4264±0.2391 \\
AttU-Net         & 0.8989±0.0185  & 5.2799±1.0656 & 1.4524±0.2388 \\
MixAttNet        & 0.8990±0.0202  & 5.5529±0.8239 & 1.5440±0.2164 \\
UENTR            & 0.8925±0.0189  & 6.5094±0.8034 & 1.8066±0.2512 \\
SwinUNETR        & 0.8976±0.0195  & 6.2951±0.7365 & 1.7283±0.2204 \\
AtlasSeg         & \textbf{0.9105±0.0163} & \textbf{5.0457±0.9324} & \textbf{1.3308±0.1935}    \\ \hline
\end{tabularx}
}
\end{table*}

\begin{table*}
\centering
\caption{Comparison of DSC for seven tissues using different methods (n=32). The best value for each measurement has been shown in boldface text.}
\label{table2}
\fontsize{9pt}{14pt}\selectfont
\resizebox{15cm}{!}{
\begin{tabularx}{15cm}{@{\hskip 0.15in}X@{\hskip 0.4in}X@{\hskip 0.15in}X@{\hskip 0.15in}X@{\hskip 0.15in}X@{\hskip 0.15in}X@{\hskip 0.15in}X@{\hskip 0.15in}X@{\hskip 0.15in}}
\hline
Network          & CSF   & GM    & WM    & Ventricles & Cerebellum & dGM   & Brainstem \\ 
\hline
U-Net            & 0.9372 & 0.9012 & 0.9417 & 0.7912 & 0.8201 & 0.9193 & 0.9133 \\
U-Net++          & 0.9364 & 0.8961 & 0.9360 & 0.8083 & 0.8517 & 0.9206 & 0.9213 \\
AttU-Net         & 0.9382 & 0.9006 & 0.9411 & 0.8183 & 0.8499 & 0.9251 & 0.9189 \\
MixAttNet        & 0.9372 & 0.9010 & 0.9426 & 0.8212 & 0.8459 & 0.9256 & 0.9193 \\
UENTR            & 0.9368 & 0.8978 & 0.9385 & 0.8016 & 0.8359 & 0.9134 & 0.9238 \\
SwinUNETR        & 0.9391 & 0.9012 & 0.9427 & 0.8180 & 0.8354 & 0.9261 & 0.9209 \\
AtlasSeg         & \textbf{0.9420} & \textbf{0.9116} & \textbf{0.9478} & \textbf{0.8470} & \textbf{0.8556} & \textbf{0.9359} & \textbf{0.9338} \\ 
\hline
\end{tabularx}
}
\end{table*}

\begin{figure*}[!t]
\centering
\includegraphics[width=\textwidth]{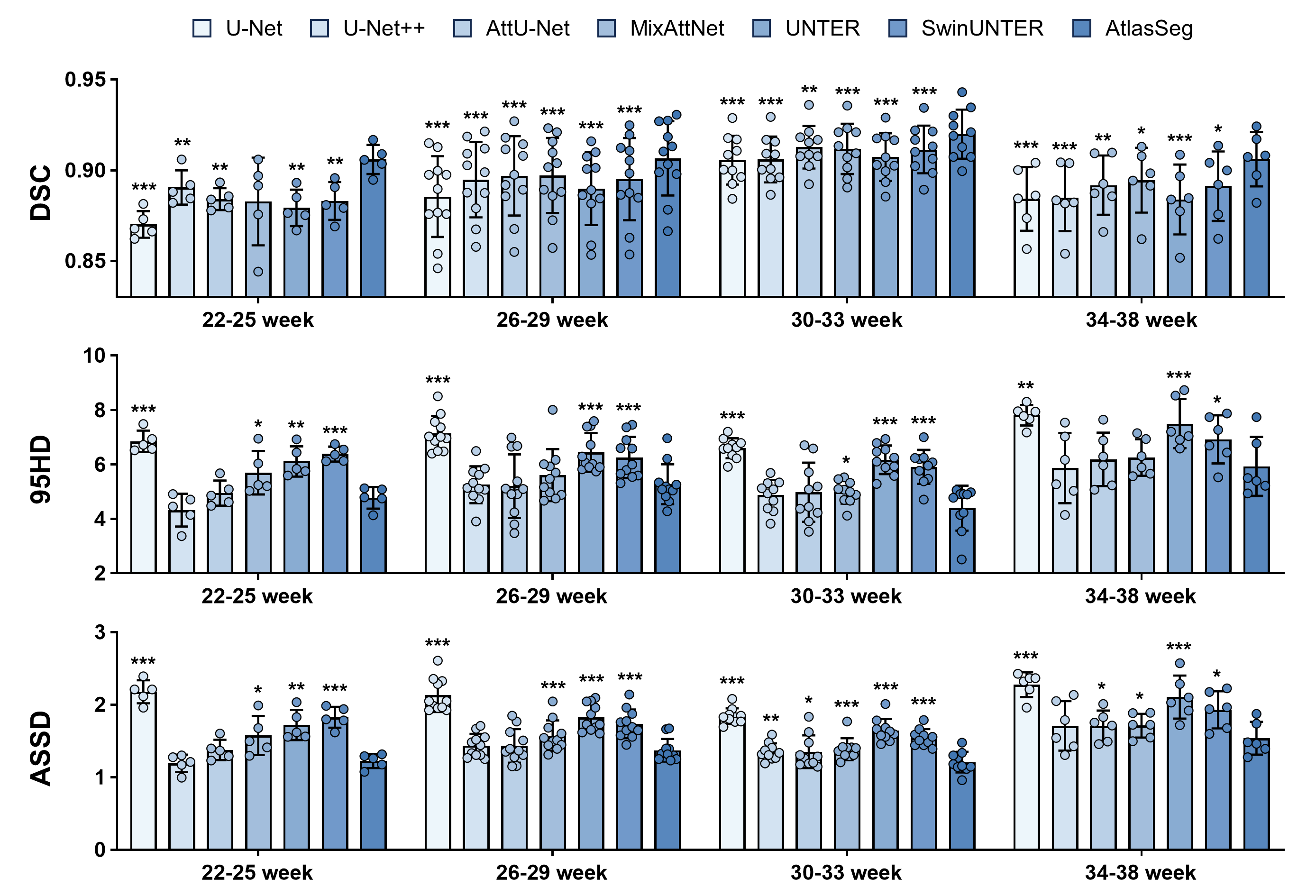}
\centering
\caption{Segmentation performance across different fetal age groups using various networks in terms of DSC, 95HD, and ASSD. Bars indicate mean values, and error bars represent the SDs. The fetal age range from 22 to 38 weeks is divided into four groups, with each containing 4 or 5 ages. The asterisk labels the significant difference (*p$<$0.05, **p$<$0.01, ***p$<$0.001) with respect to AtlasSeg using paired t-test. The performance of AtlasSeg is the most accurate in each group and more robust across groups.}
\label{fig3}
\end{figure*}

\begin{figure*}[!t]
\centering
\includegraphics[width=\textwidth]{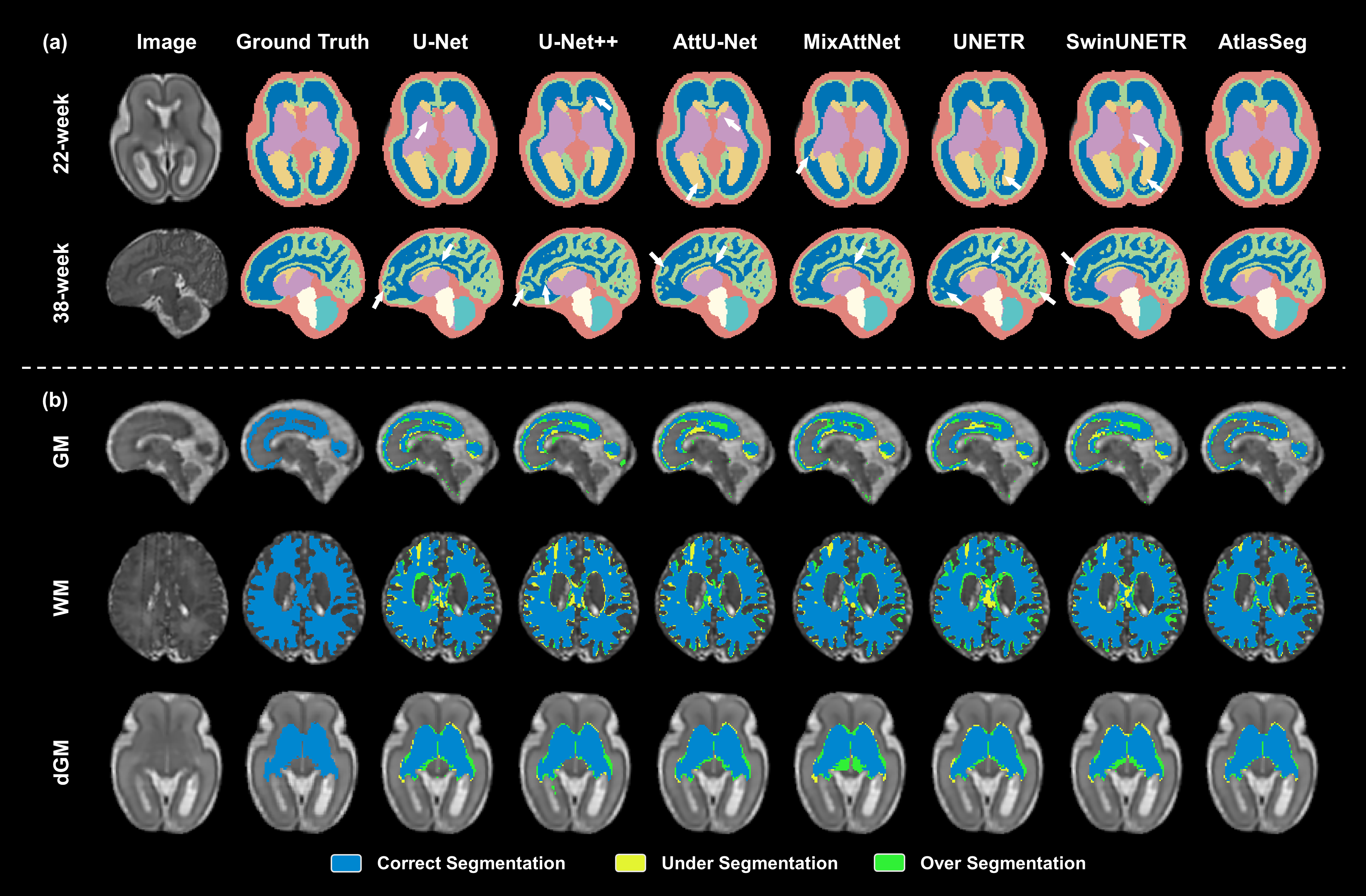}
\centering
\caption{Visual comparisons of segmentation labels predicted by different networks. (a) Multi-label predictions at the earliest (22 weeks) and latest (38 weeks) fetal ages. Incorrect predictions are highlighted with white arrows. (b) Single-label predictions, including the tissue of gray matter (GM, 23 weeks), white matter (WM, 37 weeks), and deep gray matter (dGM, 24 weeks). Correct, over, and under segmentations are indicated in blue, yellow, and green, respectively.}
\label{fig4}
\end{figure*}

\begin{figure*}[!t]
\centering
\includegraphics[width=\textwidth]{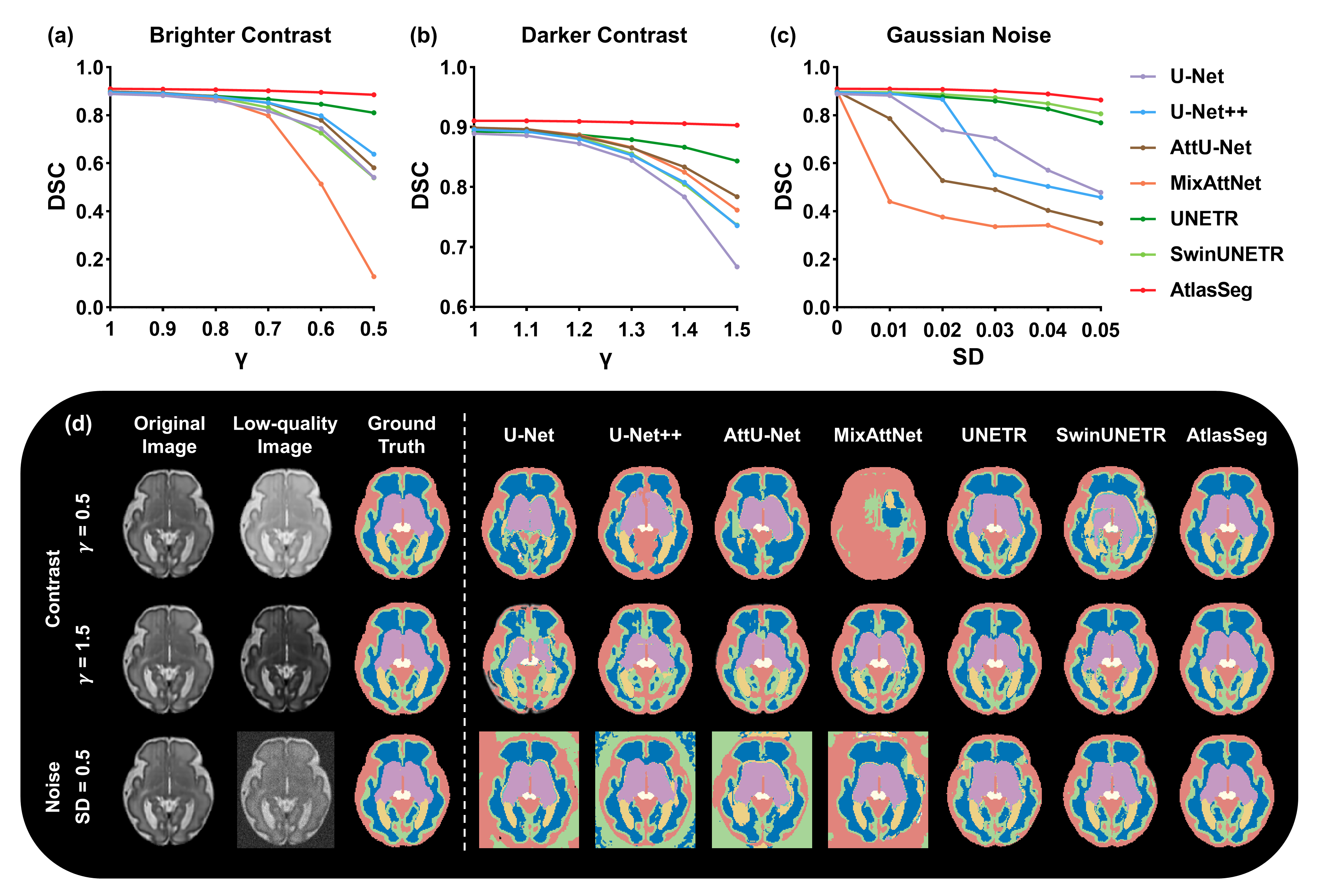}
\centering
\caption{Results of different networks on images with added perturbations. (a) and (b) involve gamma correction to adjust the image contrast, where image becomes brighter when $\gamma$$<$1 and darker when $\gamma$$>$1. (c) Gaussian noise with a standard deviation ranging from 0 to 0.05. AtlasSeg demonstrates superior robustness across all three tests. (d) Multi-label predictions of three perturbation tests on a 30-week fetal brain. The left three columns display the original image, the corrupted low-quality image, and the ground truth. The seven columns on the right show the segmentation labels generated by different networks. All six competing networks exhibit noticeable segmentation errors, while AtlasSeg maintains a significantly higher segmentation accuracy. SD = standard deviation.}
\label{fig5}
\end{figure*}

\subsection{Implementation and Training}
The network was implemented using PyTorch \cite{paszke_pytorch_nodate} and trained on an NVIDIA GeForce RTX 3090 GPU with a batch size of 4. We trained our network using the Adam optimizer \cite{kingma_adam_2017} with an initial learning rate of 1e-3 in an end-to-end manner, and reduced the learning rate by a factor of 0.9 if there was no improvement on the validation set for 5 epochs. Training stopped after 800 epochs, each comprising 25 iterations. Data augmentation of random 3D flip, rotation up to 35 degrees, contrast adjustment, and elastic deformation were used during training. All comparison and ablation networks were trained with the same datasets, augmentation techniques, and optimization hyperparameters. While MixAttNet was trained using its specific loss function, other networks used the combined CE and dice loss for training. Both UNETR and SwinUNETR were implemented using MONAI (https://monai.io/), and the feature size were set as 48.

\section{Result}
The comparison of fetal brain MRI segmentation algorithms were presented in Table \ref{table1}, based on DSC, 95HD, and ASSD measurements. AtlasSeg demonstrated superior performance, with the highest average DSC of 0.9105, the lowest average 95HD of 5.0457, and the lowest average ASSD of 1.3308. Paired t-tests showed significant differences (\(p<0.001\)) between AtlasSeg with other networks, expect for 95HD of U-Net++ and AttU-Net. AttU-Net and MixAttNet achieved suboptimal performance in DSC, even surpassing two transformer-based networks. U-Net++ exhibited suboptimal results in two distance-related measurements. Table \ref{table2} presented DSC for seven fetal brain tissues segmented by different networks. The highest DSC value for each tissue is highlighted in bold. AtlasSeg consistently outperformed all other methods, achieving the highest DSC across all tissues.

We divided the GA range of the entire test set (22-38 weeks) into four subgroups: early (22-25 weeks), mid-early (26-29 weeks), mid-late (30-33 weeks), and late (34-38 weeks), and presented the performance of different networks within each group in Fig. \ref{fig3}. Bar denoted the group means, with error bars representing SDs. The asterisk labeled the significant difference of each network with respect to AtlasSeg using paired t-test. AtlasSeg consistently outperformed competing networks in all age subgroups. Notably, while DSC improvements between AtlasSeg and other networks were marginal in mid-early ($\Delta$DSC=0.133) and mid-late ($\Delta$DSC=0.108) groups, more pronounced improvements were observed in the early ($\Delta$DSC=0.242) and late ($\Delta$DSC=0.176) groups. This highlighted the enhanced robustness of AtlasSeg under extreme GA conditions, attributable to its integration of GA prior knowledge. Further supporting its robustness across diverse ages, AtlasSeg demonstrated lower variability in mean DSC across 4 subgroups (SD=0.0059) compared to competing networks (SDs ranging from 0.0079 to 0.0126).

Fig. \ref{fig4} provided a qualitative comparison of segmentation labels generated by different networks. Fig. \ref{fig4}a displayed cases from the earliest (22 weeks) and latest (38 weeks) GAs in the test set, with segmentation errors annotated by white arrows. AtlasSeg achieved the closest results to the ground truth across both GA extremes. At 22 weeks, competing networks often mis-segmented the anterior and posterior CSF as connected structures in the mid-brain region. At 38 weeks, erroneous cortical segmentation was observed in the anterior region in several networks. Fig. \ref{fig4}b showed single-label segmentation results for three diagnostically critical tissues: GM, WM, and dGM, sampled from distinct cases. Segmentation errors, including over-segmentation and under-segmentation, were highlighted in yellow and green. In WM segmentation (second row), a striping artifact (introduced during image reconstruction) persisted in the left-anterior region of the image. This artifact caused under-segmentation in all competing networks, while AtlasSeg exhibited minimal error, attributable to its integration of shape priors.

Fig. \ref{fig5} demonstrated the results of generalizability experiments under three low-quality image conditions: increased contrast, decreased contrast, and additive Gaussian noise. In a and b, $\gamma$ represented the contrast shift parameter. Competing networks exhibited significant performance degradation with increasing contrast deviation, whereas AtlasSeg remained robust across all conditions ($\Delta$DSC$<$0.03). MixAttNet showed heightened sensitivity to increased contrast (Fig. \ref{fig5}a), while U-Net performed the worst under reduced contrast (Fig. \ref{fig5}b). In noise-corrupted experiment (Fig. \ref{fig5}c), AtlasSeg demonstrated minimal performance decline ($\Delta$DSC=0.047), whereas all four CNN-based networks suffered substantial degradation. Both transformer-based networks also showed resilience to noise. Fig. \ref{fig5}d showed the qualitative comparison. The second column displayed perturbed input images. Across all three experiments, AtlasSeg’s segmentation labels (rightmost column) aligned most closely with the ground truth. Notably, in the noise experiment (bottom row), CNN-based networks mistakenly classified noise-affected background as valid tissue labels, while transformer-based networks and AtlasSeg maintained high performance, which could be attributed to their global context modeling and shape priors, respectively.

In the ablation experiments, the impact of attention manners on AtlasSeg's performance was depicted in Table \ref{table3}. Integrating MSA into the backbone U-Net improved the DSC from 0.8891 to 0.8963, outperforming U-Net++ but remaining inferior to MixAttNet. However, the addition of atlas branch (via concatenation of feature maps from two branches) significantly enhanced the performance to 0.9070, highlighting the contribution of the atlas branch to segmentation accuracy. Further improvements were observed when feature fusions were replaced with SA and MSA. Table \ref{table4} showed the impact of attention positions. Adding attention only at the first encoding layer or the final decoding layer slightly improved performance. Variants with attention integrated within all encoding or decoding layers showed enhanced performance. AtlasSeg, with attention applied at every stage, achieved the best segmentation performance. Table \ref{table5} demonstrated the impact of channel number in the atlas branch, with optimal segmentation performance achieved at 8 channels (one-quarter of segmentation branch). One possible explanation could be insufficient channels failed to encode atlas-specific anatomical features, while excessive channels disproportionately amplified the influence of atlas priors.

\begin{table}[t]
\centering
\caption{The Ablation Performance of AtlasSeg Based on Varying Attention Manners}
\label{table3}
\fontsize{9pt}{14pt}\selectfont
\setlength{\tabcolsep}{0.5cm}
\begin{tabularx}{\columnwidth}{@{\hskip 0.05in}l@{\hskip 0.15in}c@{\hskip 0.3in}c@{\hskip 0.3in}c@{\hskip 0.33in}X}
\hline
Network & Atlas Branch & SA & MSA & DSC↑ \\
\hline
\multirow{2}{*}{UNet} & & & & 0.8891±0.0202 \\
& & & \checkmark & 0.8963±0.0207 \\
\hline
\multirow{3}{*}{AtlasSeg} & \checkmark & & & 0.9070±0.0173 \\
& \checkmark & \checkmark & & 0.9090±0.0161 \\
& \checkmark & & \checkmark & \textbf{0.9105±0.0163} \\
\hline
\end{tabularx}
\end{table}

\begin{table}[]
\centering
\caption{The Ablation Performance of AtlasSeg Based on Varying Attention Positions}
\label{table4}
\fontsize{9pt}{14pt}\selectfont
\resizebox{\columnwidth}{!}{
\begin{tabularx}{\columnwidth}{@{}l@{}*{9}{>{\centering\arraybackslash}X}l@{}}
\hline
\multirow{2}{*}{Network} & \multicolumn{9}{c}{Position of Attention}   & \multirow{2}{*}{DSC↑} \\
\cmidrule(){2-10}
                    & $E_1$ & $E_2$ & $E_3$ & $E_4$ & $B$ & $D_4$ & $D_3$ & $D_2$ & $D_1$ &  \\
\hline
Early Fusion  & \checkmark   &     &     &     &   &     &     &     &     & 0.9059±0.0173   \\
Late Fusion   &     &     &     &     &   &     &     &     & \checkmark   & 0.9048±0.0177   \\
Encoder Fusion  & \checkmark & \checkmark & \checkmark & \checkmark &   &     &     &     &     & 0.9079±0.0172         \\
Deconder Fusion &     &     &     &     &   & \checkmark & \checkmark & \checkmark & \checkmark & 0.9078±0.0166         \\
AtlasSeg     & \checkmark & \checkmark & \checkmark & \checkmark & \checkmark & \checkmark & \checkmark & \checkmark & \checkmark & \textbf{0.9105±0.0163}    \\  
\hline  
\end{tabularx}
}
\end{table}

\begin{table}[t]
\centering
\caption{Performance Comparison Based on Channel Numbers}
\label{table5}
\fontsize{9pt}{14pt}\selectfont
\setlength{\tabcolsep}{0.5cm}
\begin{tabularx}{0.3\textwidth}{@{\hskip 0.1in}l@{\hskip 0.33in}X} 
\hline
\raggedright Channel number & DSC↑ \\
\hline
4  & 0.9053±0.0173 \\
8 (AtlasSeg) & \textbf{0.9105±0.0163} \\
16 & 0.9052±0.0179 \\
32 & 0.9015±0.0184 \\
\hline
\end{tabularx}
\end{table}

\section{Discussion}
Accurate fetal brain tissue segmentation is challenged by the rapid and complex changes in brain anatomy, contrast, and tissue morphology during pregnancy. Previous works typically overlooked the GA characteristics and relied on end-to-end manners to memorize features from all training data, leading to significant performance variations across different pregnancy stages. In this work, to explicitly learn age-related features, we introduced AtlasSeg, which processed T2-weighted MRI and GA-matched atlas in parallel using a dual-U-Net architecture. By leveraging a publicly available fetal brain atlas, AtlasSeg extracted GA-specific features in the atlas branch to guide the segmentation branch producing more accurate tissue segmentation across varying GAs. Through dense multi-scale attention connections, AtlasSeg facilitated feature interactions between segmentation and supporting atlas branches. We adopted a multi-label segmentation task, where networks were trained to segment the fetal brain into eight labels: CSF, GM, WM, ventricles, cerebellum, dGM, brainstem, and background. We compared AtlasSeg with six competing networks and evaluated the segmentation performance across different GA subgroups. The results showed that AtlasSeg outperformed other networks across all GA stages, with greater improvements when dealing with extreme GA cases. 

There were several well-developed networks for medical image segmentation, such as U-Net and U-Net++. The fully convolutional operations and encoder-decoder architecture made them a benchmark in the field of medical image segmentation. Some networks had incorporated attention mechanisms to improve the segmentation performance, like AttU-Net and MixAttNet. Recently, some transformer-based network, such as UNETR and SwinUNETR, relied on self-attention mechanisms and global contextual reasoning to achieve more accurate segmentation. However, these networks had not addressed the unique characteristics in fetal brain MRI segmentation, i.e., the segmentation performance was often GA-dependent and significantly affected by the changing anatomy during development. The end-to-end training manner in the current framework did not incorporate fetal age information. Instead, they attempted to memorize the morphology of each training sample and implicitly learned the age-specific features. 

AtlasSeg was therefore proposed to explicitly encode the GA-specific information. At each GA, an atlas along with its labels were inserted to provide guidance for the segmentation in terms of label location, cortical morphology, and tissue contrast. Upon the baseline U-Net as segmentation branch, AtlasSeg constructed an atlas branch with a similar structure, including the same convolution operations, stages, and feature map sizes. The main difference was that the base number of feature map in atlas branch was set to one-quarter of that in segmentation branch to reduce model complexity. After feature extraction, how to integrate the age-specific feature from the atlas branch into the segmentation branch was the core of the network. Lots of feature fusion approaches had been proposed, such as add fusion, late concatenation, SA \cite{ferrari_cbam_2018}, channel attention \cite{hu_squeeze-and-excitation_nodate}, and self-attention \cite{zhang_self-attention_nodate}. In our work, we selected the MSA as the interaction of two features. Multi-scale convolutions with a group of kernel sizes had been employed by many networks, including Inception \cite{szegedy_going_2015}, PSPNet \cite{zhao_pyramid_2017}, and MixAttNet \cite{dou_deep_2021}. This approach proved beneficial in our study, where the size of feature being fused varied across different stages. Specifically, in the shallower stages, larger features needed a combination of large and small receptive fields to extract contextual information at multiple scales. The ablation results demonstrated the effectiveness of our used MSA. 

We compared AtlasSeg against six networks. As shown in Table \ref{table1} and Table \ref{table2}, AtlasSeg outperformed other networks in all measurements and tissues. Visual comparison of segmentation results further demonstrated AtlasSeg's best performance, with the lowest levels of over- and under-segmentations. More importantly, AtlasSeg demonstrated superior performance in both early and late GA cases, as shown in the quantitative and qualitative results. Fig. \ref{fig3} also highlighted the minimal performance variation of AtlasSeg across fetal development, addressing the accuracy degradation observed in conventional networks when processing cases with extreme fetal ages. Furthermore, in generalizability tests in Fig. \ref{fig5}, AtlasSeg exhibited significant resilience to perturbations such as contrast change and noise, which may be attributed to the shape prior provided by atlas. In summary, AtlasSeg's enhanced performance, consistency across fetal ages, and robustness to noise make it a more precise segmentation tool, particularly suitable for early detection of prenatal diseases, such as early-stage congenital cardiac anomalies.

The present work still has several limitations. First, the feature interaction module used in our work, MSA, may not seem state-of-the-art. In fact, we explored various feature fusion methods, such as cross-attention \cite{chen_crossvit_2021}, frequency-spatial fusion \cite{zhou_general_2024}, and registration fusion \cite{balakrishnan_voxelmorph_2019}, but none of these yielded better results, likely due to the limited training samples. Secondly, we built AtlasSeg based on U-Net. More experiments are needed to investigate other recent transformer-based backbones, which may further enhance performance. Moreover, the atlas branch and dense attentive connections also introduce additional network parameters and require excessive memory. Future work will focus on finding a more effective prior integration method, using multi-task learning \cite{ren_task_2020} or multi-modal language-image model \cite{radford_learning_nodate} to enhance the network’s efficiency. 

\section{Conclusion}
In this work, we presented AtlasSeg, an atlas prior guided convolutional neural network for accurate fetal brain MRI tissue segmentation. To handle fetal brains with various shapes and structures during pregnancy, AtlasSeg was designed with an incorporation of prior knowledge in the form of GA-specific atlases to provide contextual guidance. AtlasSeg had dual-U-Net architecture to process input MRI volumes and corresponding atlases, and also dense feature attentive fusion connections to facilitate a feature flow between two branches. AtlasSeg outperformed six competing networks in multi-label segmentation, and achieved more consistent performance across different fetal age groups. AtlasSeg also demonstrated greater robustness to noise and contrast changes. Given its accuracy, consistency, and robustness, AtlasSeg stands as a promising tool for reliable fetal brain MRI tissue segmentation, particularly suited for early-pregnancy diagnostic and quantitative assessments.

\section*{References}
\bibliographystyle{IEEEtran}
\bibliography{ref1.bib}

\end{document}